\documentclass[12pt,preprint]{emulateapj}

\begin{document} 
\submitted{}
\title{From Supermassive Black Holes to Dwarf Elliptical Nuclei:  a Mass Continuum}
\author{Elizabeth H. Wehner and William E. Harris} 
\affil{Department of Physics \& Astronomy, McMaster University} 
\affil{Hamilton, ON L8S 4M1, Canada}
\email{wehnere@physics.mcmaster.ca, harris@physics.mcmaster.ca} 

\shorttitle{Central Massive Objects}
\shortauthors{Wehner \& Harris}

\begin{abstract} 
Considerable evidence suggests that supermassive black holes reside at
the centers of massive galactic bulges.  At a lower galactic mass range, many dwarf
galaxies contain extremely compact nuclei that structurally resemble massive globular
clusters.  We show that both these types of central massive objects (CMO's) define a
single unbroken relation between CMO mass and the luminosity of their host galaxy
spheroid. Equivalently, $M_{CMO}$ is directly proportional to the host spheroid mass over
4 orders of magnitude.  We note that this result has been simultaneously and independently
identified by \citet{cote06}, see also \citet{ferr06}.  
We therefore suggest that the dE,N nuclei may be the low-mass
analogs of supermassive black holes, and that these two types of CMO's may have both 
developed starting from similar initial formation processes.
The overlap mass interval between the two types of CMO's is
small, and suggests that for $M_{CMO} > 10^7 M_{\odot}$, the formation of a black hole was
strongly favored, perhaps because the initial gas infall to the center was too rapid
and violent for star formation to occur efficiently.
\end{abstract}


\section{Introduction} 
\label{intro}

Even before the presence of supermassive black holes (SBH) was confirmed in galactic
centers \citep{kr95}, studies of how these dark massive objects relate to their
host galaxies were underway \citep[see][for a review]{ff05}. \citet{kr95} found that the masses of SBHs are correlated
with the dynamical masses of the galactic bulges in which they reside.  This correlation,
along with others such as the $M_{BH}-\sigma$ relation \citep{geb00,fm00} and
$M_{BH}-L_{bulge}$ \citep{korm93,md02,mh03,bettoni03} suggest that the formation of black
holes is intricately linked with galaxy evolution and the early formation of the galaxy
spheroid.

\citet{mag98} were the first to explore the $M_{BH}-M_{bulge}$ relationship in detail,
finding that black hole mass goes up essentially in direct proportion to bulge mass,
$M_{BH} \sim M_{sph}^{0.96}$. Later work by \citet{hr04} improved the precision of
previous discussions; their results are consistent with \citet{mag98}, although they find
a slightly higher slope 1.12$\pm0.06$.  They also find that the $M_{BH}-M_{bulge}$
correlation is as tight as the more often explored $M_{BH}-\sigma$ relation. 

What is less well known is whether this trend continues to lower masses.  Observational
searches for so-called intermediate-mass black holes (IBH) at the centers of smaller
galaxies have been attempted, mostly without success, all the way down to the lowest-mass
dwarf spheroidal satellites of the Milky Way {\citep[e.g.][]{val05,mac05,greene04}.  Well
established cases with $M_{BH} < 10^7 M_{\odot}$ are extremely rare, the smallest one
being the black hole in M32 at $M_{BH} = 2.5 \times 10^6 M_{\odot}$ \citep{ver02}.  It
has been suggested that the small ($\sim10^3 M_{\odot}$) black holes proposed to exist in
some massive globular clusters such as M15 and M31-G1 may represent the low-end
extrapolation of the $M_{BH} - M_{sph}$ relation \citep{geb02,gur04}.  However, even if
these GC-type black holes exist, they have formed within a stellar system of vastly
different structure and smaller scale size than a galactic spheroid.
Although galactic bulges
and GCs are both ``hot'' stellar systems, they follow quite different fundamental-plane
relations \citep[e.g.][and numerous additional papers cited there]{has05,mcl05,kis06}:
for E galaxies and bulges, the Faber-Jackson relation shows that luminosity scales with
central velocity dispersion as $L \sim \sigma^4$, while GCs follow a relation
closer to $L \sim \sigma^2$ 
(equivalent to nearly constant effective radius). The two sequences merge at the
luminosity level of the ``Ultra-Compact Dwarfs'' (UCDs; cf.~the references cited above).

However, small galaxies often contain another type of dense, massive system at their
centers. A high fraction of dwarf elliptical galaxies in particular contain well defined,
sharp nuclei: these are compact and (mostly) old stellar systems that have effective
radii of a few parsecs, typically resembling massive globular clusters in size and
structure \citep[e.g.][]{dur97,miller98,lmf04,geha02,gkp05}. Some of the most massive
known globular clusters such as NGC 6715, $\omega$ Cen, or M31-G1 are thought to be the
relic nuclei of satellite dE,N galaxies that have been tidally stripped and absorbed by
their bigger parents \citep[e.g.][]{freeman93,zinn88,lay00,maj00,mey01,mcw05}. In turn, the dE,N nuclei
themselves show no evidence for harboring massive black holes themselves \citep{geha02}
(while their models suggest that BH formation is possible, they find only a generous dE,N
black hole upper mass limit of $10^7 M_{\odot}$).

In this Letter, we explore the possibility that the dE,N nuclei may be the low-mass
analogs of the SBHs found in more massive galaxies, and that there is a single mass
continuum that connects these two types of central massive objects (CMOs).

\section{The $M_{CMO}-M_{spheroid}$ Relation}
\label{relation}

In order to explore the more general link between a galaxy's central massive object (CMO)
and its surrounding spheroidal component, we combine the relevant data in the literature
for nucleated dwarf ellipticals and for galaxy bulges with SBHs. In Figure 1, we show the
correlation between CMO masses and the luminosity of their surrounding spheroid: the dE,N
nuclei data are taken from the high-resolution HST WFPC2 imaging of \citet{lmf04}, and
the SBH data from the compilations of \citet{ff05} (SBH masses) and \citet{tr02} (spheroid data).  
(Although the SBH masses are more
commonly plotted against the bulge velocity dispersion $\sigma_{sph}$, we
cannot do this for more than a handful of the dE,N galaxies since only a few 
have measured $\sigma-$values.  We therefore compare these two types of systems 
through their spheroid luminosities, which still give a well defined sequence.)
For the dE,N nuclei, direct
dynamical masses are not yet available for most of them, so we have calculated their
masses by assuming a mass-to-light ratio appropriate for massive globular clusters,
$(M/L) \simeq 3$; within factors of two, the general trend shown in Fig.~1 is quite
insensitive to the adopted $(M/L)$ of the nucleus.

Figure 1 shows that the two types of CMOs form a single continuous trend of increasing
$M_{CMO}$ with increasing host spheroid luminosity that seamlessly connects the
dE nuclei with the SBH.  There is, also, a fairly distinct
break near $M_{CMO} \sim 10^7 M_{\odot}$, where the most luminous dE nuclei stop and the
main population of SBHs begins.  The small overlap between the two types of CMOs should
only partly be due to observational selection effects: finding SBHs with $M < 10^7
M_{\odot}$ is challenging since they generate only small and subtle dynamical effects on
their surrounding bulges; but any dE nuclei more luminous than $\sim 10^7 L_{\odot}$
would easily stand out observationally, and 
it seems plausible to conclude that they must be quite rare.  While there is 
a change in slope at or near this transition point, 
the fact that these two object types, formerly thought to be 
unrelated, form a continuous sequence, suggests an interesting connection between 
these two populations.

For the SBHs, $M_{CMO}$ scales closely in direct proportion to $L_{sph}$, as
mentioned above.  By contrast, Fig.~1 shows that the slope for the dE nuclei is
distinctly flatter: \citet{lmf04} do not quote a specific correlation, 
but a useful value is given by
\citet{gkp05}, who find from their $(B,I)$ photometry of dE,N galaxies in Virgo an
overall scaling of $L_{nuc} \sim L_{dE}^{0.7}$. However, for progressively
lower-luminosity dE galaxies, it is well known that their \emph{luminosity} becomes
increasingly less representative of their total \emph{mass}, because the dark-matter
component becomes relatively more dominant. Intrinsic scatter aside, it is plausible to
expect that the CMO mass might be most strongly determined by the depth of the potential
well that it finds itself in; that is, by the total mass enclosed by the galaxy spheroid.
A more informative way to replot the correlation in Fig.~1 might then be a graph of CMO
mass versus spheroid mass.

%
%
\begin{figure} 
\figurenum{1} 
\plotone{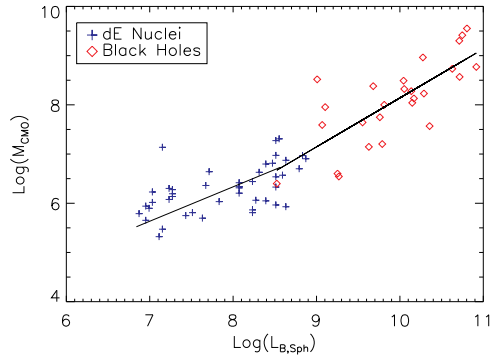} 
\caption{Log of the mass of the Central Massive Object (dE nucleus or central black hole) in 
a galaxy, $M_{CMO}/M_{\odot}$, plotted versus the log of the luminosity of the spheroid of the host
galaxy, in Solar luminosities.  Supermassive black holes are the open 
diamonds, while dE,N nuclei are the crosses.  For the black holes, $L_B$ refers to the total
bulge luminosity of the galaxy, while for the dE,N nuclei, $L_B$ refers to the luminosity
of the entire dE.  The line shown for the dE,N data represents the $B$-band slope of 0.7 
measured by \citet{gkp05}.}
\label{fig1} 
\end{figure}

In Figure 2, we show $M_{CMO}$ versus the calculated values of $M_{sph}$. For the
large-galaxy bulges, we have calculated total masses from their published magnitudes and
their mass-to-light ratio ($M/L$). 
\citet{tr02} include bulge $(M/L)$ data for most of their
galaxies, which we used to obtain $M_{sph}$ where possible.  For those without listed
$M/L$ ratios, we used $M/L = 4.0$, the midpoint of the published values.  
These $M/L$ ratios are derived from measured velocity
dispersions in the hot component of the galaxy, and thus represent the total mass, from both 
luminous and dark matter, contained within the central spheroidal component of the galaxy.
By employing the individual $M/L$ values, we implicitly take into account any global trend
of the mass-to-light ratios with galaxy luminosity; for example,
for large ellipticals, the mass-to-light ratios scale roughly as
$M/L \sim L^{0.3}$ \citep[e.g.][]{van91}. The $(M/L)$ values from \citet{tr02} agree
well with this trend.

While published $M/L$ data exist for numerous large galaxies and their spheroidal 
components, the literature for dE galaxies is not nearly so complete.  Since so few dE velocity
dispersions have been measured, we have used $M/L$ to convert their 
luminosities into masses.  According to cold dark matter theory, dwarf ellipticals 
form from average-amplitude density fluctuations in the early universe, and are dark 
matter dominated \citep{ds86}.  There have been several subsequent efforts 
to measure the relationship between a galaxy's $M/L$ ratio and its luminosity: 
If we write $(M/L) \sim L^{-\alpha}$, various recent results in the literature 
give $\alpha$ in the range $0.2 - 0.4$ \citep[e.g.][]{ds86,pc93,mf99,kf04,ddzh01}. 
We adopt a scaling $\alpha \simeq 0.3$, and
assume for the brightest dE's in the sample $(M/L) \simeq 5$ with the view that the
combined stellar population of the largest dwarf E galaxies is closely similar to that of the bulge
of a bigger galaxy. This scaling then implies that the dE spheroid mass \emph{including}
the dark matter scales as $M_{sph} \sim L^{0.7}$, which is parallel to the relation
between $M_{CMO}$ and $L$ in Fig.~1 above.  This result immediately implies $M_{CMO} \sim
M_{sph}^{1.0}$ for the dwarf galaxies.

%
%
\begin{figure} 
\figurenum{2} 
\plotone{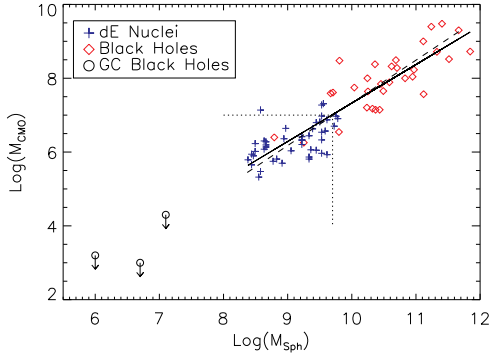} 
\caption{Log of the mass of the central massive object (dE nucleus or central black hole) versus
the log of the mass of the old spheroidal component, both in Solar masses.  The linear regression (solid)
yields a slope of 1.04, supporting the idea that the direct correlation between 
$M_{BH}-M_{bulge}$ continues to lower masses.  The average of the fits in 
X and Y is indicated by the dashed line.  The dotted line marks the transitional 
potential-well mass, below which central SBH formation is inhibited.  The globular cluster data (open circles) were taken from \citet{ku05} and references therein.
}
\label{fig2} 
\end{figure}

A straightforward linear regression fit of $M_{CMO}$ versus $M_{sph}$
to the combined data in Fig.~2 yields 

\[ {\rm log} M_{CMO} = (1.04 \pm 0.06) {\rm log} M_{sph}- (3.10 \pm 0.60) \ \ \ \ (1) \]

\noindent whereas a fit assuming the scatter to be distributed equally between
the two quantities yields the dashed line in Figure 2, given by  

\[ {\rm log} M_{CMO} = (1.18 \pm 0.10) {\rm log} M_{sph}- (4.43 \pm 0.81) \ \ \ \ (2) \]

The quoted uncertainties represent minimum $1\sigma$ errors.
These fits are mutually consistent to well within their 2$\sigma$ uncertainty.  
In summary, we find that the CMO mass in all these galaxies, defined
either as the central black hole or the stellar-system nucleus, scales accurately in
direct proportion to the surrounding mass of the spheroid over 4 orders of magnitude.  
We note that this trend has also been independently and simultaneously identified by \citet{cote06,ferr06}.

For sake of completeness, we also include in Fig.~2 the upper limits for the 
suggested black hole masses in the globular clusters $\omega$ Cen, M15, and
M31-G1 \citep[][and references therein]{ku05}.  As suggested by \citet{geb02} and 
\citet{gur04}, they lie close to the $\sim 10^3 - 10^4 M_{\odot}$ 
black hole mass regime that would be expected from an extreme
downward extrapolation of the SBH relation.  As noted above, however, it is unclear
whether they belong generically to the same sequence.

\section{Discussion}
\label{discuss}

Although the connection we propose between SBHs and dE,N nuclei rests on some indirect
argument, the unbroken continuum shown in the Figures, regardless of modest
slope changes, is highly suggestive that both
types of objects may represent two possible endpoints of
a similar initial formation process.
Both types of objects are at the deepest
points in their surrounding large-scale potential wells, and both are suggested to be
quite old; that is, they were among the first substructures to have formed within the
initial spheroid of the host galaxy \citep[e.g.][]{lmf04,ferr02}.  

We suggest that the key factor deciding whether a CMO would end up as either a supermassive
black hole or a globular-cluster-like stellar system may have been the rapidity
of the initial gas infall rate to the center.  This, in turn, will be driven by
the depth of the large-scale potential well of the young host galaxy.
For gas gravitating within the dark-matter halo of
the spheroid, \citet{hnr98} find that the mass deposition rate at the center will
increase with the halo circular velocity as $\dot M \sim v^3$, or 
equivalently \citep{ferr02} $\dot M \sim M_{DM}$.
Evidently, for the highest-mass bulges, gas falls in rapidly and
violently, favoring the formation of a black hole and quick initial growth.  It seems
reasonable then to speculate that for smaller galaxies with slower central accumulation
rates, say timescales very roughly longer than $\sim 10^6$ years, the central gas mass would have
time to make stars, halt any further inward dissipation and collapse of the gas,
and settle into a subsystem strongly resembling a
massive globular cluster or UCD.

If this speculation is essentially correct, then Figures 1 and 2 empirically identify the
transition point where the CMO changes from a dE,N nucleus to a SBH.  This mass occurs at
approximately $M_{CMO} \simeq 10^7 M_{\odot}$ or equivalently $M_{sph} \simeq 10^{10}
M_{\odot}$. Above that limit, formation of a SBH is strongly favored. Contrarily,
galactic potential wells much smaller than this appear less likely to support the
formation and growth of a central black hole; instead, the infalling gas generates a more
familiar, though still very compact, stellar system.  In this respect, the suggestions of
\citet{ferr02} that potential wells ``... of mass smaller than $\sim 5 \times 10^{11}
M_{\odot}$ are increasingly less efficient at forming SBHs -- perhaps even unable to form
them'', and of \citet{hnr98} that ``there might be a physically determined lower limit
[above $10^6 M_{\odot}$] to the mass of a supermassive black hole'' seem prescient.
On the basis of the preceding arguments,
we speculate that when it becomes possible to model this rapid central formation
process more completely, the accumulation of $\sim 10^7 M_{\odot}$ of gas within a 
``transition'' bulge mass of $\sim 5 \times 10^9 M_{\odot}$ will be found to take
$10^6$ years or less.

Our results are also more consistent with dE,N models in which the formation of the
dwarf galaxy and its massive, distinct nucleus are linked and coeval. Lastly, we point out that
this link may provide a physical reason for the long-observed upper mass limit of about
$10^7 M_{\odot}$ for old globular clusters \citep[e.g.][]{har06}. 

\acknowledgements

EHW and WEH thank the Natural Sciences and Engineering Research Council of Canada for
financial support.  We are grateful to Pat C\^ot\'e and Laura Ferrarese for comments on an early version, and for cooperating with the publication process.

\end{document}